\documentclass[aps,prb,reprint,twocolumn]{revtex4-1}

\usepackage{amssymb, amsmath, amsfonts}
\usepackage{graphicx}

\begin{document}

\title{Modern Telecommunications: \\A Playground for Physicists?}
\author{Aris L. Moustakas}
\affiliation{Department of Physics, \\National and Kapodistrian University of Athens, Greece}

\begin{abstract}
This report summarizes some of the material that was presented by the author during the 2015 Les Houches Summerschool on ``Random Matrices and Stochastic Processes''. In these Lectures, various applications of Random Matrix Theory in modern telecommunications are reviewed. The aim is to introduce the Physics community to a number of relevant problems that can be analyzed using such tools, while at the same time briefly describing the way these methods are applied. More specifically, two applications on wireless communications and two on optical communications are presented.
\end{abstract}

\maketitle


\section{Introduction}
\label{sec:Intro}

Data traffic in wireless networks has been increasing exponentially for a long time and is expected to continue this trend. The emerging data-hungry applications, such as video-on-demand and cloud computing, as well as the exploding number of smart user devices demand the introduction of disruptive technologies. An analogous situation appears in the case of wireline (mostly fiber-optical) traffic, where the currently deployed infrastructure is expected to soon reach its limits, leading to the so-called ``capacity crunch'' \cite{Tkach2010_ScalingCrunch}.

One way to counter this trend is the parallelization of information transmission in the spatial domain, thereby transmitting multiple data streams in parallel by using the same infrastructure (antennas) over the air in wireless communications or within the same optical fiber. The challenge is that, unlike the parallel use of orthogonal frequencies, the cross-talk between the different data streams can be significant, since there are no naturally occurring orthogonal modes due to the randomness of the medium. \cite{Foschini1998_BLAST1} first developed an algorithm in the context of wireless communications for multiple antennas at the transmitter and receiver that could compensate this additional interference and promise unprecedented increases in data throughput. The acronym used for this system in the engineering community is ``MIMO'', signifying multiple input and multiple output data streams.

Since it was first proposed, the technology has matured enough, at least in the context of wireless communications, so that current projections of what the next generation wireless systems will likely be envision massive (in terms of their number) antenna arrays transmitting parallel streams of data to many users nearby (hence called Massive MIMO) \cite{Andrews2014_WhatWill5GBe}. Not surprisingly, similar projections are made for the case of fiber-optical communications, where fibers with multiple cores have been proposed \cite{Morioka2012_CommMag_Enhancing_OCommsFibers}.

It is therefore important to analyze the performance of such MIMO systems in the environments they are envisioned to operate. One very useful tool in this direction has been random matrix theory, with the help of which both exact and asymptotic expressions  for various quantities of interest have been derived. After all in several of the occurring problems, such as Massive MIMO mentioned above, the asymptotic limit usually taken in random matrix theory is actually realistic. Therefore, such results are useful for performance prediction and network design, but also to provide intuition to system engineers on the way the network operates. This is so, because the obtained results show which system parameters are relevant, and which not. As a result, research in this field can be rewarding both for its scientific rigor but also for the direct applicability of its results.

The aim of these lectures is to introduce the physics community to a number of relevant problems in communications research and the types of solutions that have been used to tackle them. In the process, interested readers may be able to further acquaint themselves with research in engineering bibliography cited herein.

\subsection{Outline}
\label{sec:Outline}

After a brief introduction to basic metrics and quantities of interest in Section \ref{sec:ITBasics}, Section \ref{sec:WirelessComms} describes the solution to two problems in the context of wireless communications. More specifically, in Section \ref{sec:CapacityCorrelatedAntennas} the statistics of information capacity in wireless MIMO systems are analyzed, while Section \ref{sec:Mobility} deals with the effects of macroscopic mobility of users. Section \ref{sec:OpticalComms} provides two different ways to calculate the statistics of the mutual information in fiber-optical communications, all using various methods of random matrix theory. Generalizations, similar problems, shortcomings and open problems are also mentioned in the text.

\section{Information Theory Basics}
\label{sec:ITBasics}

In this section we introduce a few metrics that are relevant in information transmission, and will be used in further sections.

\subsection{Information Capacity}

A key quantity in information theory is the mutual information between an input random variable $X$ and an output random variable $Y$ and is defined as
\begin{equation}\label{eq:mutual-info-scalar-def}
I(X,Y) = -\iint dX dY \mbox{Pr}(X,Y) \log\left[\frac{\mbox{Pr}(Y)}{\mbox{Pr}(Y|X)}\right]
\end{equation}
where the probability distribution $\mbox{Pr}(Y|X)$ describes the type of noise the input $X$ is subjected to, in order to produce the output $Y$. The maximum of this quantity with respect to the input distribution $\mbox{Pr}(X)$, subject to certain constraints, such as maximum transmitted power, is called information capacity and represents the maximum number of nats (which are bits in the Neperian basis) that can be transmitted error-free per channel use. For a simple additive Gaussian-noise channel of the form
\begin{equation}\label{eq:IO-scalar}
  Y=\sqrt{\rho} X + Z
\end{equation}
where  $Z\sim {\mathcal CN}(0,1)$ is the noise, and $\rho$ is the signal to noise ratio, the mutual information is maximized with a Gaussian input of unit variance $X\sim {\mathcal CN}(0,1)$. In this case the capacity can be expressed as \cite{Cover_Thomas_book}
\begin{equation}\label{eq:Shannon-cap-scalar}
  I = \log\left(1+\rho\right)
\end{equation}

The above analysis can be generalized in the case of $N$ transmit and receive antennas with an average power constraint imposed at the transmitter. The corresponding channel equation can be expressed as
\begin{equation}\label{eq:IO-MIMO}
  {\bf y} = \sqrt{\frac{\rho}{N}} {\bf G} {\bf x} + {\bf z}
\end{equation}
where now ${\bf G}$ is the matrix of channel coefficients between the transmit and receive antennas and ${\bf y}$ and  ${\bf z}$ are the $N$ dimensional output signal and noise vector respectively, with the latter assumed to be independent and complex Gaussian with unit variance. In this case as well, the optimum input distribution is complex Gaussian. If $\bf G$ is known at the transmitter the input covariance matrix $E[{\bf x}{\bf x}^\dagger]$ can be optimized to take advantage of this knowledge. However, for simplicity, here we assume that this information is not available at the receiver, in which case the covariance is unity, i.e. $E[{\bf x}{\bf x}^\dagger]={\bf I}_N$. When the channel is known at the receiver then the information capacity (in nats) is
\begin{equation}\label{eq:Shannon-cap-MIMO}
  I_N = \log\det\left({\bf I}_N+\rho{\bf G}{\bf G}^\dagger\right)
\end{equation}
Note that for convenience, we have absorbed a factor of $N^{-1/2}$ in the definition of $\bf G$. This expression is also valid for channels where the transmitter has $n_t$ antennas available and the receiver has $n_r$ antennas, i.e. when $\bf G$ is $n_t\times n_r$. In this case, we need to replace ${\bf I}_N$ by ${\bf I}_{n_r}$. The capacity represents the maximum rate that can be transmitted error-free for a given channel matrix $\bf G$. Since the channel matrix is randomly distributed the capacity itself is a random quantity. Its average $E[I_N]$ provides an estimate of what kind of throughput rate one should expect on average. However, since $\bf G$ varies (albeit slowly) over time, the instantaneous rate must be fed back to the transmitter to encode the data accordingly. If this is not possible, there is always a finite probability that $\bf G$ will change in such a way that the encoded rate is not supported in the transmission and errors will occur. In this case the outage capacity is relevant, which is defined as the value $R_{\mbox{out}}$ of the cumulative distribution of $I_N$ above for which the probability that $I_N<R$ is $p_{\mbox{out}}$, i.e.
\begin{equation}\label{eq:outage_cap-MIMO_def}
  p_{\mbox{out}} = \mbox{Pr}(I_N<R_{\mbox{out}})
\end{equation}
Therefore, the full distribution of $I_N$ is important to characterize the transmission performance.

\subsection{Linear Precoders}

In the previous subsection we described the performance of a system of transmit and receive antenna arrays, assuming that the received signal from the antennas can be jointly processed. Often however the receive antennas are not collocated as they correspond to different mobile users communicating with a multi-antenna base-station. This is the typical situation in a so-called massive MIMO system. In this case the information capacity of each user takes the form of \eqref{eq:Shannon-cap-scalar} with $\rho$ substituted by an appropriately defined signal-to-ratio. However, in this case there is significant interference between users. One way to counter this is to pre-multiply the signal vector at the transmitter with an appropriately chosen matrix $\bf V$. Due to the linearity of matrix multiplication, this approach is called linear precoding. As a result, the received signal at user $k=1,\ldots,n_r$ can be expressed as
\begin{equation}\label{eq:linear_preq_def}
  y_k = {\bf g}_k^T {\bf Vx} + \sigma  z_k
\end{equation}
where ${\bf g}_k^T$ is the $k$th row of the matrix $\bf G$. Clearly, this only makes sense if the transmitter has some information about the $\bf G$. There are several forms of precoding matrices, one of which is  the so-called ``zero-forcing'' precoding matrix, which amounts to the pseudo-inverse of $\bf G$, i.e.
\begin{equation}\label{eq:ZF_def}
  {\bf V} = {\bf G}^\dagger \left({\bf GG}^\dagger\right)^{-1} {\bf P}^{1/2}
\end{equation}
where $\bf P$ is a diagonal matrix with elements the designated receive powers of each user $p_k$. Clearly, this matrix exists only if $n_t\leq n_r$. The benefit of using this precoding matrix is that the signal at each receiver is completely decoupled. Indeed plugging \eqref{eq:ZF_def} into \eqref{eq:linear_preq_def} results to the trivial
\begin{equation}\label{eq:linear_preqZF}
  y_k = \sqrt{p_k} x_k + \sigma z_k
\end{equation}
and therefore the signal-to-noise ratio requirements are immediately met if $p_k=\rho_k*\sigma^2$, where $\rho_k$ is the requested signal-to-noise ratio. The price for this is the increased transmitted power, which can be evaluated to be
\begin{equation}\label{eq:total_power}
  P_{\mbox{tot}} =\frac{1}{n_t} \mbox{Tr}\left[\left({\bf GG}^\dagger\right)^{-1} {\bf P}\right]
\end{equation}
Additional precoding techniques exist in the literature \cite{Sanguinetti2014_EnergyConsumptionMIMOMobility}, which tend to trade between interference cancellation at the receiver end and power consumption or channel information at the transmitter. When more than one antennas exist in the receiver, similar techniques can be applied there as well. However, in all cases one is left with an object, such as in \eqref{eq:total_power}, which depends on the channel randomness. Hence once again, random matrix theory can be of immediate help to get quantitative estimates.

\section{Wireless Communications: Replicas and Mobility}
\label{sec:WirelessComms}

In this section we will provide two specific applications of random matrix theory in wireless communications.

Before moving ahead, it is important to introduce the statistics of the propagation channel matrix $\bf G$. A good and reliable model for its elements $G_{i\alpha}$ is that they are complex Gaussian random variables due to multiple scattering. The correlations of the matrix elements can be evaluated in the diffusion approximation to be \cite{Moustakas2000_BLAST1}
\begin{equation}\label{eq:GG*_corr}
  E\left[G_{i\alpha}G_{j\beta}^*\right] = \frac{\rho }{n_t} R_{ij} T_{\alpha\beta}
\end{equation}
In the above equation, $R_{ij}$ and $T_{\alpha\beta}$ are the elements of the correlation matrices between the antennas at the receiver and transmitter arrays, respectively. $R_{ij}$ can be expressed as \cite{Moustakas2000_BLAST1}
\begin{equation}\label{eq:R_corr_def_chi}
  R_{ij} = \ell\left({\bf r}\right) \int d\Omega_{\bf k}\,  \chi_i({\bf k})\, \chi_j({\bf k})\, e^{i {\bf k} {\bf d}_{ij}}\, w({\bf k})
\end{equation}
where $\chi_i({\bf k})$ is the response of the antenna $i$ at incoming wavevector ${\bf k}$, ${\bf d}_{ij} $ the vector between antennas $i$ and $j$ and $w({\bf k})$ the weight of incoming power with a similar expression for $T_{\alpha\beta}$ (without the $\ell(\cdot)$-term). Thus antennas are more decorrelated the further apart compared to the wavelength they are and the more evenly over angles the incoming (or relevant outgoing) power is spread. Also, $\ell(\cdot)$ is the average power loss due to propagation and $R_{tr}$ is the distance between receiver and transmitter array. A typical model for $\ell(x)$ is $\ell(x)=|x|^{-\beta}$, where the pathloss exponent is usually taken to be $\beta=4-5$ \cite{Calcev2004_3GPP_SCM}. Also, we have included the factor $1/n_t$ here that was absorbed into $\bf G$ earlier.

\subsection{Capacity of Correlated Antennas}
\label{sec:CapacityCorrelatedAntennas}

In his section we  introduce a method based on replicas to obtain the asymptotic moments of the capacity distribution in the large antenna limit assuming the above discussed channel model. This methodology was first developed by \cite{Sengupta2000_BLAST1} and extended in \cite{Moustakas2000_BLAST1,Moustakas2003_MIMO1}. While not rigorous it provides results in a few number of steps, which took a while to be established rigorously \cite{Hachem2006_GaussianCapacityKroneckerProduct}.

The starting point is the moment generating function
\begin{equation}
\label{eq:g(mu)_wireless_MIMO_def}
  g(-\mu) = E\left[e^{-\mu I_N}\right] = E\left[\det\left({\bf I}_{n_r} + \rho {\bf G G}^\dagger\right)^{-\mu}\right]
\end{equation}
The key trick in the calculation is to express the determinant above as a Gaussian complex integral, so that the matrices $\bf G$ will appear in the exponent and can then averaged over. After some algebra we obtain
\begin{align}
\label{eq:g(mu)_wireless_MIMO1}
  g(-\mu) = \iint d{\bf X} d{\bf Y} &\,e^{-\frac{1}{2}\mbox{Tr}\left[{\bf X}^\dagger{\bf X} + {\bf Y}^\dagger{\bf Y} \right]} \\ \nonumber
  & \times E\left[e^{-\frac{\sqrt{\rho}}{2}\mbox{Tr}\left[{\bf X}^\dagger{\bf G}{\bf Y} - {\bf Y}^\dagger{\bf G}^\dagger{\bf X}\right] } \right]
\end{align}
where $\bf X$ and $\bf Y$ are $n_r\times \mu$ and $n_t\times\mu$ dimensional complex matrices with the appropriate integration measure $d\bf X$ and $d\bf Y$, respectively. After averaging over $\bf G$, we obtain:
\begin{equation}
\label{eq:g(mu)_wireless_MIMO2}
  g(-\mu) = \iint d{\bf X}d{\bf Y} e^{-\frac{1}{2}\mbox{Tr}\left[{\bf X}^\dagger{\bf X} + {\bf Y}^\dagger{\bf Y} + \frac{\rho}{2n_t}{\bf X}^\dagger{\bf R}{\bf X}{\bf Y}^\dagger{\bf T}{\bf Y}\right]}
\end{equation}
The quartic term in the exponent cannot be integrated as such. However, in the large $n_t$ limit we can treat it in a mean-field way. We now introduce the $\mu\times\mu$ matrices $\mathcal T$ and $\mathcal R$ through the following identity
\begin{eqnarray}
\label{eq:g(mu)_wireless_MIMO3}
  1 &=& \int D{\mathcal T} \, \delta\left({\mathcal T}-\frac{\rho}{2\sqrt{n_t}}{\bf X}^\dagger {\bf R} {\bf X}\right) \\ \nonumber
  &=& \iint D{\mathcal T}D{\mathcal R}\, e^{\mbox{Tr}\left[{\mathcal RT} - \frac{\rho}{2\sqrt{n_t}}{\mathcal R}{\bf X}^\dagger {\bf RX}\right]}
\end{eqnarray}
where the $\delta$-function appearing above is shorthand for a product of $\delta$-functions on all real and imaginary parts of the elements of the matrix ${\mathcal T}$. The integration of the elements of ${\mathcal T}$ is over the real axis, while that for the elements of ${\mathcal R}$ are over the imaginary axis, in agreement with Fourier integration. We then insert this identity into \eqref{eq:g(mu)_wireless_MIMO2} getting
\begin{eqnarray}
\label{eq:g(mu)_wireless_MIMO4}
  g(-\mu) &=& \iint D{\mathcal T}D{\mathcal R} \, e^{\mbox{Tr}\left[{\mathcal TR}\right]} \iint d{\bf X}d{\bf Y} \\ \nonumber 
   && \times e^{-\frac{1}{2}\mbox{Tr}\left[{\bf X}^\dagger{\bf X} + {\bf Y}^\dagger{\bf Y} + \frac{\rho}{\sqrt{n_t}}{\bf X}^\dagger{\bf R}{\bf X}{\mathcal R} + \frac{1}{\sqrt{n_t}}{\bf Y}^\dagger{\bf T}{\bf Y}{\mathcal T}\right]}
\end{eqnarray}
which, after integrating over $\bf X$, $\bf Y$ reduces to
\begin{eqnarray}
\label{eq:g(mu)_wireless_MIMO5}
  g(-\mu) &=& \iint D{\mathcal T}D{\mathcal R} \, e^{-{\mathcal S}}  \nonumber \\
  {\mathcal S} &=& \log\det\left[{\bf I}_{n_t}\otimes{\bf I}_{\mu} + \frac{1}{\sqrt{n_t}}  {\bf T}\otimes{\mathcal T} \right] \\ \nonumber
  &+& \log\det\left[{\bf I}_{n_r}\otimes{\bf I}_{\mu} + \frac{\rho}{\sqrt{n_t}}  {\bf R}\otimes{\mathcal R} \right] - \mbox{Tr}\left[{\mathcal TR}\right]
\end{eqnarray}
The remaining integrals over the elements of the matrices ${\mathcal T}$, ${\mathcal R}$ will be performed using the saddle-point method. To do so, we need to ``guess'' the structure of these matrices at the saddle point. In the usual replica literature, the dynamic degrees of freedom (e.g. spin variables) take discrete values. Hence the corresponding correlation matrices at the replica symmetric saddle-point need to be symmetric over permutations over the replica indices. In contrast, here the dynamic variables, i.e. $\bf X$, $\bf Y$ are continuous and thus have $U(\mu)$ rotational symmetry in replica space. Therefore, at the replica-symmetric saddle-point, ${\mathcal T}$ and ${\mathcal R}$ need to be scalars, which we express them as
\begin{eqnarray}
\label{eq:saddle point1}
  {\mathcal T} &=& t\sqrt{n_t} \, {\bf I}_\mu +\delta{\mathcal T}  \\ \nonumber
  {\mathcal R} &=& r\sqrt{n_t} \, {\bf I}_\mu +\delta{\mathcal R}
\end{eqnarray}
Plugging these expressions into \eqref{eq:g(mu)_wireless_MIMO5} we get to leading order
\begin{eqnarray}
  g(-\mu) \approx e^{-{\mathcal S}_0}
\end{eqnarray}
where
\begin{eqnarray}
  {\mathcal S}_0 \equiv \mu \Gamma_0 &=& \mu\left( \log\det\left[{\bf I}_{n_t} +   t {\bf T} \right] \right.\\ \nonumber
  &+& \left.\log\det\left[{\bf I}_{n_r} + \rho  r {\bf R}\right] - n_t rt\right) 
\label{eq:g(mu)_wireless_MIMO6}
\end{eqnarray}
with $r,t$ satisfying the saddle-point equations
\begin{eqnarray}
\label{eq:saddle point2}
  r &=& \frac{1}{n_t} \mbox{Tr}\left[\frac{{\bf T}}{{\bf I}_{n_t} + t{\bf T}}\right]  \\ \nonumber
  t &=& \frac{1}{n_t} \mbox{Tr}\left[\frac{\rho {\bf R}}{{\bf I}_{n_r} + \rho r{\bf R}}\right]
\end{eqnarray}
Since $g'(0)=-E[I_N]$, we immediately see, that to leading order in $n_t$, $E[I_{n_t}]=\Gamma_0$.

To obtain higher moments of the distribution, we need to expand ${\mathcal S}$ in powers of $\delta {\mathcal R}$ and $\delta {\mathcal T}$. At the saddle point, the linear terms vanish, hence the leading term is the quadratic one,
\begin{eqnarray}
\label{eq:g(mu)_wireless_MIMO7}
  {\mathcal S} = {\mathcal S}_0 &-& \frac{1}{2}\sum_{\mu,\nu} \left[\delta {\mathcal R}_{\mu\nu},\,\,\delta {\mathcal T}_{\mu\nu}\right]
  \left(
    \begin{array}{cc}
      r_2 & 1 \\
      1 & t_2 \\
    \end{array}
  \right)
  \left[
  \begin{array}{c}
    \delta {\mathcal R}_{\nu\mu} \\
    \delta {\mathcal T}_{\nu\mu}
  \end{array}
  \right] \\ \nonumber 
  &+& {\mathcal O}\left(\delta {\mathcal R}^3,\delta {\mathcal T}^3\right)
\end{eqnarray}
where
\begin{eqnarray}
\label{eq:saddle point3}
  r_2 &=& \frac{1}{n_t} \mbox{Tr}\left[\frac{{\bf T}^2}{\left({\bf I}_{n_t} + t{\bf T}\right)^2}\right]  \\ \nonumber
  t_2 &=& \frac{1}{n_t} \mbox{Tr}\left[\frac{\rho^2 {\bf R}^2}{\left({\bf I}_{n_r} + \rho r{\bf R}\right)^2}\right]
\end{eqnarray}
Integrating over the quadratic term in the exponent by appropriately rotating the contour of integration close to the saddle point, we obtain
\begin{eqnarray}
\label{eq:g(mu)_wireless_MIMO8}
  g(-\mu) \approx e^{-\mu \Gamma_0 -\frac{\mu^2}{2}\log\left(1-r_2t_2\right)}
\end{eqnarray}
As a result, the variance of the mutual information takes the following simple form
\begin{eqnarray}
\label{eq:g(mu)_wireless_MIMO9}
  \mbox{Var}(I_N) = -\log\left(1-r_2t_2\right)
\end{eqnarray}
It is worth contrasting this result with the standard central limit theorem for the sum of $N$ random variables, in which the mean and the variance of the sum is ${\mathcal O}(N)$. Here the mean is ${\mathcal O}(N)$, while the variance is ${\mathcal O}(1)$. The underlying reason of these vastly reduced fluctuations can be understood by the fact that the underlying ${\mathcal O}(N)$ random degrees of freedom, i.e the eigenvalues of the matrix ${bf GG}^\dagger$  are highly correlated and (as we shall see in Section \ref{sec:Tails}) they are constrained to be located very closely in eigenvalue space.

If we continue the perturbation expansion by including cubic and quartic terms in $\delta {\mathcal T},\delta {\mathcal R}$, we obtain a ${\mathcal O}(1/n_t)$ correction term to $E[I_{n_t}]$ and a skewness of the same order \cite{Moustakas2003_MIMO1}. In fact, it can be established that all higher moments vanish when $n_t,n_r\to\infty$, thereby making the distribution asymptotically Gaussian. Interestingly, it can also be shown that the replica-symmetric saddle-point is stable \cite{Moustakas2007_MIMO1}.

\begin{figure}[tb]
\begin{center}
\includegraphics[width=1.1\columnwidth]{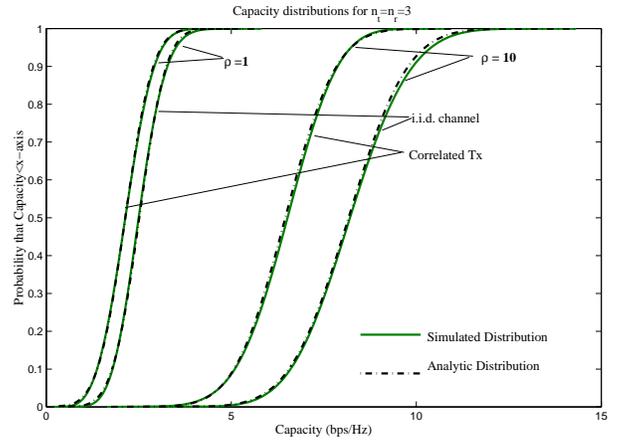}
\end{center}
\caption{Cumulative probability distributions of the capacity for a system with $n_t=n_r=3$ antennas. In one set of curves have uncorrelated antennas, while the  transmitter antennas of the other are $d=\lambda$ apart with a $\delta=5^o$ angle-spread. Two different values of $\rho$ are used. The theoretical curves are Gaussian distributions, with mean and variance calculated as above. The agreement with the simulated curves is quite good. }
\label{fig:CDF_WMIMO}
\end{figure}

One reason these results are quite useful is that they are applicable not only for the case of very large antenna numbers, but also for just a few antennas. This can be seen explicitly in Fig. \ref{fig:CDF_WMIMO}, where the agreement with simulations is remarkable even for $n_t=3$. In conclusion, we have seen a first example, where random matrix theory can provide useful results in wireless communications.

\subsection{Effect of Mobility on Energy Consumption}
\label{sec:Mobility}

In addition to traffic growth, another related big challenge is the increasing energy consumption of cellular infrastructure equipment. As a result, energy consumption has to be a key ingredient in the design of future cellular networks, especially in new rural regions of the developing world, where the electrical grid is unreliable or even non-existing. In this section we will analyse the distribution of energy consumption for a particular case of linear precoding discussed in Section \ref{sec:ITBasics} when we take the mobility of users into account. Once again, the large system size will simplify  the analysis considerably.

We consider a base-station (BS) with $n_t$ antennas serving $n_r$ mobile single antenna users with $c\equiv n_r/n_t<1$, in a square region centered at the BS with side length $L$. Here we focus in the downlink case, where the BS acts as a transmitter. In order to guarantee a certain signal to noise ratio $\rho_k$ to user $k$, the transmitting array precodes the signal using the Zero-Forcing precoding matrix appearing in \eqref{eq:ZF_def}. As a result, the total transmitted power is given by \eqref{eq:total_power}. This power is time-dependent due to the movement of the users through the temporal variation of the channel coefficients. This in turn has two components. One originates form the relatively slow variation of the pathloss due to the macroscopic movement of the users. The characteristic time for this is $\sim L/v$, where $v$ is the typical velocity of the users. There is another much faster variation of $\bf G$ due to multiple (or Rayleigh) scattering. In the engineering literature, these fluctuations are called ``fast-fading''. The timescale here is much shorter, $\sim \lambda/v$. The analysis below will distinguish between these two processes.

For concreteness, we employ a simple mobility model for users, namely that of a Brownian motion, which is the continuous version of a simple random walk. Hence, $\mbox{Pr}({\bf x},{\bf x}';t-t')$, the probability of a user to be at position $\bf x$ at time $t$ given that he was at $x'$ at time $t'$ satisfies the diffusion equation
\begin{equation}\label{eq:diff_equation}
\frac{\partial \mbox{Pr}({\bf x},{\bf x}';t-t')}{\partial t} = D\nabla^2 \mbox{Pr}({\bf x},{\bf x}';t-t')
\end{equation}
where the diffusion constant $D$ characterizes the small scale mobility of the user. Further, we assume periodic boundary conditions at the borders of the square to mimic the existence of other users in neighboring cells entering the current cell.

The total energy consumed by the base-station over time $T$ is given by
\begin{equation}\label{eq:Energy_Tdef}
  E_T = \int_0^T P(t) dt
\end{equation}
The aim of this section is to calculate the statistics of this quantity. We start by averaging the power $P(t)$ over fast-fading keeping the positions of the users (roughly) fixed. This step can be done by using the results of the previous section. Starting from \eqref{eq:total_power} we redefine the channel matrix ${\bf GP}^{-1/2}\to {\bf G}$ hence redefining the receiver correlation matrix ${\bf RP}^{-1}\to {\bf R}$. Then we observe that the expression in \eqref{eq:total_power} can obtained from \eqref{eq:g(mu)_wireless_MIMO6} by taking the $\rho\to\infty$ limit. More concretely,
\begin{equation}\label{eq:P(t)_defined as limit of logdet}
  P(t) = \lim_{\rho\to\infty} \left(-\rho^2\frac{d}{d\rho}\left[\log\det\left[{\bf I}_{n_r} \rho^{-1}+ \rho {\bf GG}^\dagger\right]\right]\right)
\end{equation}
Plugging in \eqref{eq:g(mu)_wireless_MIMO6} into the above equation (and re-introducing the matrix $\bf P$) gives
\begin{eqnarray}\label{eq:Phat(t)}
  {\hat P}(t) &\equiv& E\left[P(t)|\{ {\bf x}_k\}\right] \nonumber \\
   &=&  \frac{1}{n_t r} \mbox{Tr}\left[{\bf R}^{-1}{\bf P}\right] \nonumber \\
   &=&  \frac{\sigma^2}{n_t r} \sum_{k=1}^{n_r} \frac{\rho_k}{\ell({\bf x}_k(t))}
\end{eqnarray}
The last line results from the fact that since the receive antennas are so far apart, they are uncorrelated, i.e. $R_{ij}=\ell({\bf x})\delta_{ij}$. $r$ can be found in this limit to be the solution of
\begin{equation}\label{eq:r-solution-saddle-ZF}
  \frac{1}{n_t} \sum_{j=1}^{n_t} \frac{T_j}{r+cT_j} = 1
\end{equation}
where $T_j$ are the eigenvalues of the BS antenna correlation matrix $\bf T$. We see that when the transmitter antennas are uncorrelated $r=1-c$.

Now, ${\hat P}(t)$ is  time dependent only due to the macroscopic movement of the mobile users ${\bf x}_k(t)$. Averaging over their movements as well, we obtain
\begin{eqnarray}\label{eq:P_ave}
  {\overline P} = \frac{\sigma^2 c}{r} E\left[\ell^{-1}({\bf x})\right] \frac{1}{n_r}\sum_{k=1}^{n_r} \rho_k
\end{eqnarray}
Since the long-time spatial distribution of the Brownian motion is uniform the expectation above is over the whole square. As a result,
\begin{eqnarray}\label{eq:ET_ave}
  {\overline E}_T = \frac{\sigma^2 cT}{r} E\left[\ell^{-1}({\bf x})\right] \frac{1}{n_r}\sum_{k=1}^{n_r} \rho_k
\end{eqnarray}

To calculate the fluctuations of the consumed energy at the transmitter, we separate them in two parts according to their corresponding time-scales as discussed above. Hence
\begin{eqnarray}\label{eq:aveR_2parts}
  P(t)-{\overline P} = \left(P(t)-{\hat P}(t)\right) + \left({\hat P}(t) - {\overline P} \right)
\end{eqnarray}
The first part has fluctuations due to fast-fading, while the second due to user mobility. It can be observed from the relation with the previous section that  the fluctuations of the first part scale as $1/n_r^2$. This has been rigorously established in \cite{Bai2004_CLT_covariance_matrices}. Hence since the decorrelation time is $\sim\lambda/v$ we conclude that the variance of the energy due to fast-fading will be $\sim \frac{T\lambda/v}{n_r^2}{\overline P}^2$. In contrast, as we shall see, the fluctuations of ${\hat P}(t) - {\overline P} $ are of order ${\mathcal O}\left(n_r^{-1}\right)$. This is so because ${\hat P}(t)$, see \eqref{eq:Phat(t)}, has $n_r$ independent degrees of freedom (the positions of the $n_r$ mobiles users). Since the decorrelation time of the Brownian motion is $\sim L^2/D$, the variance of this energy term will be of the order $\sim \frac{TL^2}{n_rD} {\overline P}^2$.

Indeed, we can express the variance of the energy as
\begin{eqnarray}
  \mbox{Var}(E_T) =\frac{1}{2} \int_0^T dt\int_0^t \int d{\bf x}\int d{\bf x}' \ell^{-1}({\bf x}) \ell^{-1}({\bf x}')&&\\ \nonumber
   \times \left(\mbox{Pr}({\bf x},{\bf x}';t-t')-\mbox{Pr}({\bf x},{\bf x}';t+t')\right) &&
\end{eqnarray}
which, after expressing the diffusion probabilities in terms of the eigenfunctions of the diffusion equation $\Psi_{\bf n}({\bf x}=e^{i{\bf k}_{{\bf n}}{\bf x}}$ with eigenvalues ${\bf k}_{{\bf n}}=2\pi {\bf n}/L$, where ${\bf n}\in Z^2$, we obtain
\begin{eqnarray}\label{eq:VarET}
  \mbox{Var}(E_T) &=&\frac{TL^2}{8\pi^2D} \sum_{{\bf n}} \frac{\theta_{{\bf n}}^2}{|{\bf n}|^2} \\ \nonumber
  \theta_{{\bf n}}&=& \int_{-1/2}^{1/2}dx\int_{-1/2}^{1/2}dy\, \ell^{-1}\left({\bf x}L\right)\, e^{i2\pi{\bf n}{\bf x}}
\end{eqnarray}
Since $\theta_{{\bf n}}$ falls off fast with increasing $|\bf n|$, the summation in the above equation converges fast and only a few terms are necessary to evaluate it.

All higher moments of the energy can be easily shown to vanish faster in the large $n_r$ limit. Hence, the energy consumption becomes a Gaussian variable with mean and variance calculated above. This model can  be used to approximate the probability that a battery-powered BS runs out of energy and also
to design the cell radius for minimizing the energy consumption per unit area \cite{Sanguinetti2014_EnergyConsumptionMIMOMobility}.

\subsection{Discussion}
\label{sec:Wireless_discussion}

In this section we briefly discuss various generalizations of the above results.

 The results presented in with Section \ref{sec:CapacityCorrelatedAntennas} have been generalized to the calculation of the statistics of the capacity in cases where the channel matrix is not Gaussian. It turns out that only the second moment of the distribution is relevant \cite{Couillet2011_CapacityAnalysisMIMO}, at least for the mean capacity. Also, the methodology can be applied to situations where the interference itself is a random matrix. In this case, the interference channel appears in two logariths, which have different sign. In such a case, one needs to rely on supersymmetric methods, introducing integrals over Grassman variables \cite{Moustakas2003_MIMO1,Taricco2008_MIMOCorrelatedCapacity}. This results have still not been proved with more rigorous methods. Another generalization deals with the case, where the input distribution is binary rather than Gaussian \cite{Muller2003_RandomMatrixMIMO_binary}, in which case the replica approach is perhaps the only one that can provide an answer.

The random matrix analysis of the behavior of precoders as in Section \ref{sec:Mobility} is currently an active topic of research, due their possible application in next-generation communications, see for example \cite{Wagner2012_LargeSystemAnalysisLinearPrecodingMISO}. Also, since precoders usually involve matrix inversions \eqref{eq:ZF_def}, a series of works has analyzed the robust representation of precoders in terms of matrix polynomials \cite{Muller2001_PolynomialPrecoders}. Finally, note that precoders don't necessarily need to be linear, and optimization over the nonlinear precoders has also been analyzed using replicas and random matrix theory \cite{Zaidel2013_VectorPrecodingRSB}.

\section{Optical Communications: Moments and Tails}
\label{sec:OpticalComms}

One way to increase the data throughput though optical fibers is to use more channels in each fiber. At a first level one can use more than one electromagnetic mode through existing fibers \cite{Hsu06_Capacityenhancement,Tarighat2007_FundamentalsMultimodeFibers}. At a later stage, engineers envision a new generation of optical fibers, specially designed with multiple cores in each of them \cite{Morioka2012_CommMag_Enhancing_OCommsFibers,Winzer2011_OpticalMIMOCapacity}. Due to twisting, bending as well as non-linear coupling, these propagation channels mix strongly, especially when the fiber length extends over long distances. In contrast, backscattering can be assumed to be negligible. Another important difference from free space propagation in wireless communications is that fiber optical transmission is characterized by low loss. Hence the appropriate metric to describe the propagation is the transmission coefficients of the scattering matrix, since there is no reflection.

Although the total scattering matrix should be symmetric ${\bf S}={\bf S}^T$ due to time reversal symmetry \cite{Beenakker1997_MesoscopicReview}, the transmission matrix ${\bf U}$ itself does not have any other symmetries or constraints, apart from the normalization condition ${\bf U}{\bf U}^\dagger={\bf I}_N$, which is a direct consequence of the unitarity of ${\bf S}$. Therefore, in the strong mixing limit, we may neglect any bias between the various modes or cores or inhomogeneity in the mixing and assume that $\bf U$ is Haar random. It is convenient to define the channel matrix as ${\bf G} = {\bf T}^{1/2}{\bf U} {\bf T}^1/2$, where ${\bf R}$ and ${\bf T}$ are the correlations matrices of the transmitted and received signal, respectively. For example, in the optical fiber case they correspond to reflections and losses at the two edges of the link.

As a result, the corresponding MIMO channel for this system reads
\begin{equation}
 {\bf y} = \mathbf{U} {\bf x} + {\bf z}
 \label{system}
\end{equation}
with coherent detection and channel state information only at the receiver \cite{Foschini1998_BLAST1,Telatar1995_BLAST1}. ${\bf x}$, ${\bf y}$ and ${\bf z}$ are the $N$-dimensional input, output signal vectors and  unit variance noise vector, respectively, all assumed for simplicity to be complex Gaussian. We also assume no differential delays between channels, which effectively leads to frequency flat fading \cite{Winzer2011_OpticalMIMOCapacity} and no mode-dependent loss. As a result, the mutual information (in nats) can be expressed as
\begin{eqnarray}
I_{N}({\bf U})=\log \det(I + \rho {\mathbf{G}^{\dagger} \mathbf{G}})
\end{eqnarray}

As in the case of wireless communications, the above expression can be generalized to cases, where the number of active transmitters is $n_t\leq N$ and correspondingly the number of receiving elements is $n_r\leq N$. This can be done by making the matrices $\bf T$ and $\bf R$ be of rank $n_t$ and $n_r$ respectively. This may correspond to the situation, where not all transmitting or receiving channels may be available to a given link.

\subsection{Character Expansions in Communications}
\label{sec:Character}

In this section we will calculate in closed form the moment generating function of the mutual information
\begin{equation}\label{eq:mgf_opt_char}
  g(\mu) = E\left[\det\left({\bf I}_N + \rho {\bf G}^\dagger{\bf G}\right)^\mu\right]
\end{equation}
where the expectation is over the channel matrix ${\bf G}$.  To make progress, one could expand the quantity inside the expectation in \eqref{eq:mgf_opt_char} in terms of products of the matrices ${\bf U}$ and ${\bf U}^\dagger$ and then average the resulting products over the unitary group. These averages are however quite complicated and in most cases can only be treated in an asymptotic fashion \cite{Brouwer1996_DiagrammaticMethodUnitaryRMT,Argaman1996_DiagrammaticUnitaryRMT}. Instead here we employ  a different approach first introduced by Balantekin \cite{Balantekin2000_CharacterExpansionsETC}. Here, the expansion is performed using characters of the irreducible representations of the unitary group $U(N)$, the unitary matrices of size $N$. For completeness we summarize below some basic facts on representation theory of groups.  The interested reader can refer to several textbooks, including \cite{Sternberg_GroupTheory_book}.

A unitary representation $V$ of a group $G$ is a homomorphism from $G$  to $U(N)$. An irreducible representation has no non-trivial invariant subspaces. The irreducible representations of the unitary group $U(N)$ \cite{Schlittgen2003_GeneralizationsOfSomeIntegralsOverUnitaryGroups,Hua_book_GroupTheory} can be parameterized by an $N$-dimensional vector ${\bf m}=(m_1,m_2,\ldots,m_N)$, with integers $m_1\geq m_2\geq\ldots \geq m_N\geq 0$. The dimension
$d_{{\bf m}}$ of an irreducible representation is the dimension of its invariant subspace. For the case of $U(N)$ it can be shown \cite{Simon2004_CapacityOfCorrRandomWishartMatrices} that $d_{{\bf m}}$ is given by
\begin{equation}\label{eq:d_r_definition}
  d_{{\bf m}} = \left[\prod_{i=1}^N \frac{1}{(N-i)!}\right] \left(-1\right)^{\frac{N(N-1)}{2}}
  \Delta({\bf k})
\end{equation}
where $\Delta(\cdot)$ represents the Vandermonde determinant, defined as
\begin{equation}\label{eq:Vandermonde_def}
\Delta({\bf x}) \equiv \det\left(x_i^{j-1}\right) = \prod_{i>j} \left(x_i-x_j\right)
\end{equation}
and the vector ${\bf k}$ has elements
\begin{equation}\label{eq:vector_k_def}
  k_i = m_i -i + N
\end{equation}
where $i=1,\ldots,N$ and $m_i$ are the elements of the representation vector ${\bf m}$. Now, the character $\chi(g)$ of a group element $g$ in the representation $V$ is equal to the trace of the corresponding matrix, i.e. $\chi(g) = Tr\left[V(g)\right]$. Thus a character of a reducible representation can be written as a sum of characters of irreducible representations. Clearly, $\chi(g)$ depends only on the eigenvalues of $V(g)$. Calculating the characters of irreducible representations is greatly facilitated by Weyl's character formula \cite{Weyl_book}\cite{Hua_book_GroupTheory}, which for  $U(N)$ takes the form:
\begin{equation}\label{eq:Weyl_formula_def}
\chi_{{\bf m}}({\bf A}) =
\frac{\det\left(a_i^{m_j+N-j}\right)}{\Delta(a_1,\ldots,a_N)}
\end{equation}
where the index ${\bf m}$ denotes the irreducible representation  $(m_1,\ldots,m_N)$ and $a_i$, for $i=1,\ldots,N$, are the eigenvalues of ${\bf A}$ in the fundamental ($N$-dimensional) representation. For example, the characters of the one-dimensional unitary group $U(1)$ are given by $\chi_n = e^{in\phi}$, where the character index $n$ takes values $0,1,\ldots$ and $e^{i\phi}$ is the (eigen)value of an arbitrary one-dimensional matrix in $U(1)$. Thus a Fourier expansion can be seen as  an expansion in the characters of $U(1)$ group. This suggests that the characters of a group can form a good basis of expanding functions, which are
invariant under $U(N)$ group operations.

\cite{Balantekin2000_CharacterExpansionsETC} showed that the product of any function $f(x)$ of the eigenvalues $\{a_i\}$ of an invertible matrix ${\bf A}$ can be expressed in the following form
\begin{equation}\label{eq:character_expansion}
\prod_{i=1}^N f(a_i) = \sum_{{\bf m}} \alpha_{{\bf m}} \chi_{{\bf m}}({\bf A})
\end{equation}
In the above expression, $\chi_{{\bf m}}({\bf  A})$ is the character of ${\bf  A}$ in the representation ${\bf m}$ and the sum is over all irreducible representations of $U(N)$ parameterized with the vector ${\bf m}=(m_1,m_2,\ldots,m_N)$, with integers $m_1\geq m_2\geq\ldots \geq m_N\geq 0$. In \eqref{eq:character_expansion} the coefficient for each character, $\alpha_{{\bf m}}$ is given by
\begin{equation}\label{eq:alpha_x_definition}
  \alpha_{{\bf m}}= \det\left(f_{m_j+i-j}\right)
\end{equation}
where $f_n$ is the coefficient of the Taylor expansion of $f(x)$, i.e.
\begin{equation}\label{eq:f_n_definition}
  f(x)= \sum_{n=0}^\infty f_n x^n
\end{equation}
In the particular case of \eqref{eq:mgf_opt_char} ${\bf A}={\bf RUTU}^\dagger$ and $f(x)=(1+\rho x)^\mu$ so that
\begin{equation}\label{eq:f_n_expression}
  f_n= \rho^n \frac{\Gamma(\mu+1)}{\Gamma(\mu-n+1)\Gamma(n+1)}
\end{equation}
Note that, as expected, for integer $\mu$, $f_n$ vanishes for $n>\mu$. As a result, we have
\begin{equation}\label{eq:g_mu_character_expansion}
g(\mu) = \sum_{{\bf m}} \alpha_{{\bf m}} E\left[\chi_{{\bf m}}\left({\bf RUTU}^\dagger\right)\right]
\end{equation}
The expectation of the character $\chi_{{\bf m}}\left({\bf RUTU}^\dagger\right)= \mbox{Tr}\left[ {\bf R}^{{\bf m}}{\bf U}^{{\bf m}} {\bf T}^{{\bf m}} {\bf U}^{{\bf m}\dagger} \right]$, where ${\bf U}^{{\bf m}}$ etc., is the group element of ${\bf U}$ in the representation ${\bf m}$, can be obtained using the following identity \cite{Sternberg_GroupTheory_book}
\begin{equation}\label{eq:def_integration_over_U_U_bar}
  \int d{\bf U}\, U^{({\bf m})}_{ij}\, U^{({\bf m}')*}_{kl}=\frac{1}{d_{{\bf m}} }\delta_{{\bf m} {\bf m}'}
  \delta_{ik}\delta_{jl}
\end{equation}
where $d{\bf U}$ is the standard Haar integration measure and $d_{{\bf m}}$ is the dimension of the representation given above. As a result we have
\begin{equation}\label{eq:g_mu_character_expansion2}
g(\mu) = \sum_{{\bf m}} \frac{\alpha_{{\bf m}}}{d_{{\bf m}}} \chi_{{\bf m}}\left({\bf R}\right) \chi_{{\bf m}}\left({\bf T}\right)
\end{equation}
The above expression can become more transparent by using the characters of ${\bf R}$ and ${\bf T}$ through the Weyl character formula so that
\begin{equation}\label{eq:g_mu_character_expansion3}
g(\mu) = \sum_{{\bf m}} \frac{\alpha_{{\bf m}}}{d_{{\bf m}}} \frac{\det\left(t_i^{m_j+N-j}\right)}{\Delta({\bf t})} \frac{\det\left(r_i^{m_j+N-j}\right)}{\Delta({\bf r})}
\end{equation}
where $t_i$ and $r_i$ are the eigenvalues of the matrices ${\bf T}$ and ${\bf R}$, respectively.

We now need to massage the expression of $\alpha_{{\bf m}}$. To do so we start by expressing it in the following form
\begin{eqnarray}\label{eq:alpha_x_massage1}
  \alpha_{{\bf m}}&=& \prod_{i=1}^N \left(\frac{\Gamma(\mu+1)\rho^{k_i+i-N}}{\Gamma(\mu+N-k_i)\Gamma(k_i+1)}\right) \\ \nonumber &\times& \det\left[\frac{\Gamma(\mu-k_i+N)\Gamma(k_i+1)}{\Gamma(\mu-k_i+N-j)\Gamma(k_i+j-N+1)}\right]
\end{eqnarray}
where the indices ${\bf k}$ and ${\bf m}$ are related through \eqref{eq:vector_k_def}. We observe that the $(i, j)$ element of the matrix inside the square brackets above is a $(N-1)$-degree polynomial of $k_i$, indexed by the row $j=1,\ldots,N$, expressed for compactness as $\pi_j(k_i)$. By performing linear operations on the columns of the matrix we can express the determinant of the matrix as $\Delta({\bf k})$, up to an overall multiplicative factor ${\overline C}_N$, independent of $k_i$, which may be obtained by various ways, and will be discussed at the end. The key point of this calculation is that $\alpha_{{\bf m}}$ is proportional to $\Delta({\bf k})$, which can then cancel the same factor appearing in $d_{{\bf m}}$. Thus, we have
\begin{align}\label{eq:g_mu_character_expansion4}
g(\mu) = C_N \rho^{-\frac{N(N-1)}{2}}  \sum_{{\bf k}}  \prod_{i=1}^N &\left(\frac{\Gamma(\mu+1)\rho^{k_i+i-N}}{\Gamma(\mu+N-k_i)\Gamma(k_i+1)}\right) 
 \nonumber \\
&\times \frac{\det\left(t_i^{k_j}\right)}{\Delta({\bf t})} \frac{\det\left(r_i^{k_j}\right)}{\Delta({\bf r})} 
\end{align}
where we have absorbed all constant factors in $C_N$. The final step consists of using the Cauchy-Binet formula to sum over the indices $k_i$ and obtain
\begin{equation}\label{eq:g_mu_character_expansion5}
g(\mu) = C_N \rho^{-\frac{N(N-1)}{2}} \,\, \frac{\det\left[\left(1+\rho t_i r_j\right)^{\mu+N-1} \right]}{\Delta({\bf r})\Delta({\bf t})}
\end{equation}
To obtain the constant $C_N$, we may take the limit of $r_j\to 0$, successively for $j=1,\ldots,N$. Since both numerator and denominator vanish when we do this, we need to carefully apply the l'Hospital rule at each step \cite{Simon2004_CapacityOfCorrRandomWishartMatrices}. After some algebra we find
\begin{equation}\label{eq:g_mu_character_expansion6}
C_N =\prod_{i=1}^{N-1} \left(\frac{\Gamma(i+1)\Gamma(\mu+N-i)}{\Gamma(\mu+N)}\right)
\end{equation}

From the above expression, one can readily obtain the probability distribution of the mutual information $I_N(\rho)$, by Fourier transformation, i.e.
\begin{eqnarray}\label{eq:PDF_MI_optical_finite}
 \mbox{Pr}(R) = \int_{-\infty}^{\infty} \frac{d\mu}{2\pi} \, e^{-i \mu R} g(\mu)
\end{eqnarray}
In addition, the moments of the distribution can be evaluated by taking the appropriate derivatives with respect to $\mu$. For example, $E[I_N]=g'(0)$, $Var(I_N) = (\log g(0))''$, etc. For $N=2$, one obtains
\begin{align}\label{eq:ave_MI_optical_finite}
 &E[I_N]  =  -1 \\ \nonumber 
 &+ \frac{\left(\log(1+\rho a_1 b_1)+\log(1+\rho a_2 b_2)\right)(1+\rho a_1 b_1)(1+\rho a_2 b_2)}{\rho(a_1-a_2)(b_1-b_2)} \\ \nonumber
  &- \frac{\left(\log(1+\rho a_1 b_2)+\log(1+\rho a_2 b_1)\right)(1+\rho a_1 b_2)(1+\rho a_2 b_1)}{\rho(a_1-a_2)(b_1-b_2)} 
\end{align}
Clearly, these expressions become unappealing for larger $N$ and for higher moments due to the determinantal structure of $g(\mu)$. In the next section, we will show how the distribution of the mutual information can be evaluated asymptotically for large $N$ for a special form of the matrices ${\bf R}$ and ${\bf T}$. A similar expression has been derived for correlated Gaussian channels \cite{Simon2004_EigenvalueDensityOfCorrRandomWishartMatrices}.

\subsection{Tails of the mutual information}
\label{sec:Tails}

The previous section dealt with the exact calculation of the moment generating function of the mutual information. However, in real communications systems a more important metric is the probability distribution of the mutual information itself and in particular its tails, which quantify the probability of error in decoding a packet that has been sent with too high a coding rate. This is particularly true in fiber-optical communications, where very low error rates are desirable, since no feedback is available to allow the transmitter to retransmit the packet, as is the case in wireless communications.

In this section we will calculate the distribution of the mutual information of the optical MIMO channel in the large channel number $N$ regime.  By large $N$, we will signify that all $n_t,n_r,N$ go to infinity, but with fixed ratios. To be able to do this calculation, the correlation matrices at the receiver and transmitter will be take to have a simplified structure, namely ${\bf R}={\bf P}_{n_r}$ and ${\bf T}={\bf P}_{n_t}$ where ${\bf P}_n$ is the $N\times N$ projection matrix on an $n$-dimensional subspace. This simplification corresponds to idealized receiver and transmitter structures, with $n_t$ transmitter channels, $n_r$ receiver channels and several untapped channels, which may used by other transceivers or simply correspond to energy loss \cite{Simon2006_UnitaryPaper}. Since the exact nature of the subspaces will be irrelevant due to rotational symmetry, we take the matrices to be of the form ${\bf P}_n = \mbox{diag}([1,\ldots,1,0,\ldots,0])$, where the diagonal has $n$ ones and $N-n$ zeroes. In this case the joint probability distribution of the eigenvalues of the matrix ${\bf P}_{n_r}{\bf U}{\bf P}_{n_t}{\bf U}^\dagger{\bf P}_{n_r}$ has been shown to be \cite{Simon2006_UnitaryPaper,Dar2012_JacobiMIMOChannel} for $n_t+n_r\leq N$ and $n_t>n_r$
\begin{eqnarray}\label{eq:JPDF_unitary channel}
 \mbox{Pr}({\bf \lambda}) \propto \Delta({\bf \lambda})^2 \prod_{k=1}^{n_r} \lambda_k^{n_t-n_r} (1-\lambda_k)^{N-n_t-n_r}
\end{eqnarray}
while the remaining $N-n_r$ eigenvalues are zero, while for $n_t+n_r<N$ \cite{Dar2012_JacobiMIMOChannel} there are $n_t+n_r-N$ eigenvalues equal to unity, $N-n_r$ zero eigenvalues, while the remaining eigenvalues have the following density
\begin{eqnarray}\label{eq:JPDF_unitary channel2}
 \mbox{Pr}({\bf \lambda}) \propto \Delta({\bf \lambda})^2 \prod_{k=1}^{N-n_t} \lambda_k^{n_t-n_r} (1-\lambda_k)^{n_t+n_r-N}
\end{eqnarray}
The situation $n_t<n_r$ can be obtained directly from the above by interchanging $n_t, n_r$. For simplicity, we will now assume that $n_t+n_r< N$ and $n_t>n_r$.

Now, based on the above expressions, we can readily apply the methodology of the previous section to obtain the moment generating function of the mutual information, which can be expressed directly as
\begin{equation}\label{eq:MI_optical}
  I_N = \sum_{k=1}^{n_r} \log(1+\rho\lambda_k)
\end{equation}
However, given the explicit expression of the probability distribution of the eigenvalues, more can be accomplished. We will follow closely the methodology by \cite{Majumdar2006_LesHouches,Vivo2007_LargeDeviationsWishart,Dean2008_ExtremeValueStatisticsEigsGaussianRMT,Karadimitrakis2014_Optical_MIMO_Outage}
 based on the analogy to a Coulomb gas pioneered by \cite{Dyson1962_DysonGas}. The key insight is that for large $N$ the eigenvalues coalesce to a fluid that can be described as a density given by
\begin{equation}\label{eq:n(x)_def}
  n(x)=\frac{1}{n_r} \sum_{k=1}^{n_r} \delta(x-\lambda_k)
\end{equation}
We start by conjecturing that the support of $n(x)$ is compact with borders $0>a>b>1$, which we will check in the end to be the case. Hence the logarithm of the distribution function in \eqref{eq:JPDF_unitary channel} can be expressed as
\begin{eqnarray}\label{eq:logPr}
\frac{\log \mbox{Pr}(\cdot)}{n_r^2} &=&\int_a^b n(x) \left(\beta \log(1-x)+\alpha \log(x)\right) d x  \nonumber \\
&+& \iint_a^b n(x)n(y)\log{|x-y|}dy dx
\end{eqnarray}
where $\alpha=(n_t-n_r)/n_r$ and $\beta = (N-n_t-n_r)/n_r$.

To evaluate the probability density of $I_N(\rho)$ we express first in the form
\begin{eqnarray}
\mbox{Pr}(r) &=& n_r E\left[\delta\left(n_r^2\left\{ r-\int n(x) \log(1+\rho x)dx\right\}\right)\right] \nonumber \\
&=& n_r \int \frac{dk}{2\pi i} E\left[e^{n_r^2 k\left(r-\int n(x)\log(1+\rho x)\right)}\right]
\label{eq:p(r)}
\end{eqnarray}
To ensure the density $n(x)$ is properly normalized to unity, it is convenient to add another constraint in the form of a Fourier integral  as above. As a result, we obtain \begin{eqnarray}\label{eq:p(r)2}
\mbox{Pr}(r) \propto E\left[e^{-n_r^2 {\mathcal S}[n] }\right]
\end{eqnarray}
where the expectation is over all positive functions $n(x)$ and the corresponding constraint integrals mentioned above, and
\begin{eqnarray}
{\mathcal S}[n] &=& -\int n(x) \left(\beta \log(1-x)+\alpha \log(x)\right) d x \nonumber \\
 &-& \iint n(x)n(y)\log{|x-y|}dy dx \nonumber \\ 
&-& k\left(\int n(x)\log(1+\rho x)dx -r\right) \nonumber \\ 
&-& c\left(\int n(x) dx -1\right)
\label{eq:S[n]}
\end{eqnarray}
In the large $N$ limit, the path-integral is dominated by the contribution around its saddle-point(s) of ${\mathcal S}[n]$. Convexity arguments for ${\mathcal S}[n]$ can assure that any solution will be unique \cite{BenArous1997_LDWignerLaw,HiaiPetz1998_LargeDeviationsWishartEigenvalues}. Taking the functional derivative on ${\mathcal S}[n]$ with respect to $n(x)$ and setting the result to zero we obtain
\begin{eqnarray}\label{eq:saddlept}
2\int_a^b n(y)\log{|x-y|}dy &=& -\beta \log(1-x)\nonumber \\
&-&\alpha \log(x) -k\log(1+\rho x)- c
\end{eqnarray}
It is convenient to differentiate this expression with respect to $x$, which gives
\begin{eqnarray}\label{eq:saddlept1}
2{\mathcal P}\int_a^b \frac{n(y)}{x-y}\,dy = -\frac{\beta}{1-x}-\frac{\alpha}{x} -\frac{k\rho}{1+\rho x}
\end{eqnarray}
where ${\mathcal P}$ indicates the Cauchy principal value of the integral. The above equation has an appealing physical meaning, namely the balance of forces between the inter-eigenvalue (intercharge) repulsions and the forces imposed by external (one-body) potentials. Hence, the solution to this equation will provide the  equilibrium (or most probable) density of eigenvalues consistent with rate $r$. Since both $\alpha,\beta>0$, we expect an infinite force acting the charge density if either $a=0$ or $b=1$. Therefore, neither of this can be the case. Thankfully, this integral equation can be solved (see \cite{Tricomi_book_IntegralEquations,Majumdar2006_LesHouches}) with a general solution of the form
\begin{eqnarray}
\label{eq:gen_solution_int_eq0}
    n(x) = \frac{\frac{\beta\sqrt{(1-a)(1-b)}}{1-x} - \frac{k\sqrt{(1+a\rho)(1+b\rho)}}{1+\rho x} -\frac{\alpha\sqrt{ab}}{x} + C}{2\pi\sqrt{(x-a)(b-x)}}
\end{eqnarray}
where $C$ is a constant. Assuming continuity at the boundary of the support, i.e. $n(a)=n(b)=0$ we obtain
\begin{eqnarray}
\label{eq:optimal_n(x)}
n^*(x)&=&\frac{\sqrt{(x-a)(b-x)}}{2\pi (1+\rho x)}\nonumber \\
&\times&\bigg(\frac{\beta(\rho +1)}{(1-x)\sqrt{(1-a)(1-b)}}+\frac{\alpha}{x\sqrt{a b}}\bigg)
\end{eqnarray}
with the additional constraint
\begin{equation}\label{eq:p(b)constraint}
  \frac{\beta}{\sqrt{(1-a)(1-b)}}=\frac{\alpha}{\sqrt{ab}}+\frac{k\rho}{\sqrt{(1+\rho a)(1+\rho b)}}
\end{equation}
The parameters $a,b,k$ can be evaluated uniquely from the above equation, in addition to the normalization constraint
\begin{equation}\label{eq:norm_constraint}
 \int_a^b n^*(x) dx=1
\end{equation}
which demands that
\begin{equation}\label{eq:norm_condition_a>0b<1}
  \beta+\alpha+2+k=\frac{\alpha}{\sqrt{ab}}+\frac{k(1+\rho)}{\sqrt{(1+\rho a)(1+\rho b)}}
\end{equation}
and the rate constraint
\begin{eqnarray}\label{eq:r_eq}
r&=& \int n^*(x)\log(1+\rho x) \\ \nonumber
&=&  \log{\Delta\rho}+
\frac{\beta}{2\sqrt{{\bar a}_c{\bar b}_c}}\bigg[G\left({\bar a}_z,{\bar a}_z\right)-G\left({\bar a}_z,-{\bar a}_c\right)\bigg] \\ \nonumber
&+&\frac{\alpha}{2\sqrt{{\bar a} {\bar b}}}\bigg[G\left({\bar a}_z,{\bar a}\right)-G\left({\bar a}_z,{\bar a}_z\right)\bigg]
\end{eqnarray}
where $\Delta=b-a$, $z=\frac{1}{\rho}$. In the above equation, we have defined ${\bar a}=a/\Delta$, ${a_c} = 1-a$, ${\bar a}_c={a_c}/\Delta$, ${\bar a}_z=(a+z)/\Delta$ and ${\bar b}=b/\Delta$, ${\bar b}_c={b_c}/\Delta=(1-b)/\Delta$, ${\bar b}_z=(b+z)/\Delta$. The function $G(x,y)$ is given by \cite{Karadimitrakis2014_Optical_MIMO_Outage}
\begin{eqnarray}
&&G(x,y)=\frac{1}{\pi}\int_{0}^{1}\sqrt{t(1-t)}\frac{\log(t+x)}{t+y}d t\\ \nonumber
&=&-2\mbox{sgn}(y)\sqrt{|y(1+y)|}\log \left[\frac{\sqrt{x|1+y|}+\sqrt{|y|(1+x)}}{\sqrt{|1+y|}+\sqrt{|y|}}\right]\nonumber\\
&+&(1+2 y)\log\left[\frac{\sqrt{1+x}+\sqrt{x}}{2}\right] -\frac{1}{2}\left(\sqrt{1+x}-\sqrt{x}\right)^{2}\nonumber
\end{eqnarray}
The probability density in \eqref{eq:optimal_n(x)} represents the most probable distribution of eigenvalues in the subspace where $I_N=n_r r$.
We may now plug in the above expression of $n^*(x)$ into \eqref{eq:S[n]} and obtain an expression for $\mathcal{S}$ as follows
\begin{eqnarray}
\mathcal{S}^*(r)&=&\frac{k}{2}\left(r-\log\left(1+b\rho\right)\right)-\frac{\log\Delta}{2}\left(\beta+\alpha+2 \right)\nonumber \\
&-&\frac{\beta}{2}\log{{b_c}}-\frac{\alpha}{2}\log{b} \nonumber\\
&-&\frac{\beta^2}{4\sqrt{{\bar a}_c{\bar b}_c}}\left(G({\bar b}_c,{\bar b}_c)-G({\bar b}_c,-{\bar b}_z)\right)\nonumber\\
&+&\frac{\beta\alpha}{4\sqrt{{\bar a}{\bar b}}}\left(G({\bar b}_c,-{\bar b})-G({\bar b}_c,-{\bar b}_z)\right)\nonumber \\
&+&\frac{\beta\alpha}{4\sqrt{{\bar a}_c{\bar b}_c}}\left(G({\bar a},-{\bar a}_c)-G({\bar a},{\bar a}_z)\right)\nonumber\\
&-&\frac{\alpha^2}{4\sqrt{{\bar a}{\bar b}}} \left(G({\bar a},{\bar a})-G({\bar a},{\bar a}_z)\right) \nonumber \\
&-& \frac{\beta}{2\sqrt{{\bar a}_c{\bar b}_c}} \left(G(0,{\bar b}_c)-G(0,-{\bar b}_z)\right) \nonumber \\
&+& \frac{\alpha}{2\sqrt{{\bar a}{\bar b}}}\left(G(0,-{\bar b})-G(0,-{\bar b}_z)\right)
\end{eqnarray}
As a result, we have for large $N$
\begin{eqnarray}
\mbox{Pr}(r)\propto e^{-n_r^2 {\mathcal S}^*(r)}
\end{eqnarray}
To find the normalization constant, we may just divide the above expression in the absence of any constraints on $n(x)$ in \eqref{eq:p(r)}, which corresponds to $k=0$. In this case, the constraint for the mutual information is relaxed and the corresponding expression in \eqref{eq:optimal_n(x)} corresponds to the most probable distribution of eigenvalues in \eqref{eq:JPDF_unitary channel}. After some work it is easy to see that the density of eigenvalues takes a similar form to the Marcenko-Pastur equation
\begin{equation}\label{eq:MP_0}
  n_0(x) = \frac{\sqrt{(x-a_0)(b_0-x)}}{2\pi x(1-x)}
\end{equation}
where
\begin{equation}\label{eq:a0_b0_def}
a_0,b_0=\frac{\left(\sqrt{1+\beta} \pm \sqrt{(\alpha+1)(\beta+\alpha+1)}\right)^2}{\beta+2+\alpha}
\end{equation}
which has been obtained using other methods in \cite{Simon2006_UnitaryPaper,Debbah2003_UnitaryAsymptoticallyFreeMatrices}. The corresponding value of ${\mathcal S}_0$ can obtained directly by using this expression to evaluate ${\mathcal S}$ above. Analyzing the behavior of ${\mathcal S}(r)$ close  to the value of $k\approx 0$ it can be shown that
\begin{equation}\label{eq:S-S0}
{\mathcal S}^*(r)-{\mathcal S}_0 \approx  \frac{\left(r-r_{\mbox{erg}}\right)^2}{2 v_{\mbox{erg} }}
\end{equation}
where $r_{\mbox{erg}}$ is the rate obtained using $n_0(x)$ in \eqref{eq:r_eq}, which corresponds to the average (ergodic) value of the rate in the large $N$ limit and
\begin{eqnarray}\label{eq:verg}
v_{erg} =\log{\frac{(\sqrt{1+\rho b_0}+\sqrt{1+\rho a_0})^2}{4\sqrt{1+\rho b_0}\sqrt{1+\rho a_0} }}
\end{eqnarray}
where $a_0$, $b_0$ are given in \eqref{eq:a0_b0_def}.
Since the bulk of the probability distribution will be around the value $r=r_{\mbox{erg}}$, the normalization is to leading order identical to a Gaussian distribution centered at $r_{\mbox{erg}}$ with variance $v_{\mbox{erg}}$. Hence,
\begin{eqnarray}
\mbox{Pr}(r)\approx \frac{e^{-n_r^2 ({\mathcal S}^*(r)-{\mathcal S}_0)}}{\sqrt{2\pi v_{\mbox{erg}}}}
\end{eqnarray}

\begin{figure}[tb]
\begin{center}
\includegraphics[width=1.1\columnwidth]{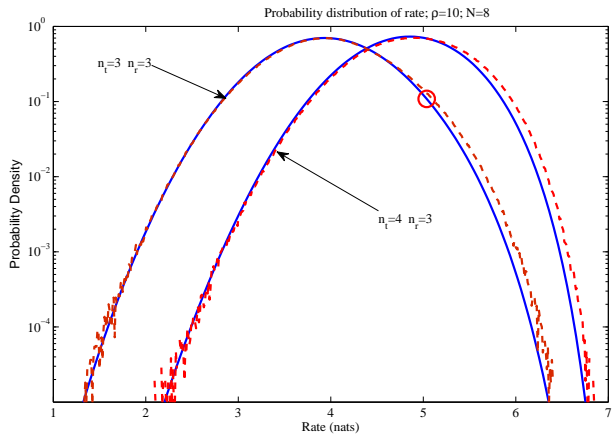}
\end{center}
\caption{Probability density of the mutual information for two different values of $n_t$. The agreement with Monte Carlo simulations is quite good. The red circle represents the point where the minimum limit of the support of the optimum distribution becomes positive $a>0$.}
\label{fig:Pout_optical}
\end{figure}

A few remarks are in order for the calculations performed above. First, although the large $N$ limit was taken here, the results are valid also for reasonably valued $n_t$, $n_r$ etc. Indeed, in Fig. \ref{fig:Pout_optical} the cumulative probability density is plotted as a function of $\rho$ for a number of representative values of $n_t$, $n_r$. As we see the agreement between Monte-Carlo simulations and this approach is pretty good.

Furthermore, it should be noted that here we only analyzed the generic case, when $\alpha,\beta>0$. When $\alpha=0$, two possible solutions for the eigenvalue density $n^*(x)$ may occur, depending on the value of $r$. The first extends all the way to the border $x=0$, with a square root singularity, while the second has a positive lower limit of its support, i.e. $a>0$. However, for a given value of $r$ only one solution is acceptable, since the other becomes negative. At some critical value of the rate $r$ there is a transition between these two solutions (see Fig. \ref{fig:Pout_optical}). Interestingly, only the third derivative of ${\mathcal S}^*(r)$ is discontinuous at this point. Similar behavior can be seen in the case when $\beta=0$ at the upper limit of the support of $n^*(x)$. This behavior has been observed in other situations \cite{Vivo2007_LargeDeviationsMaxEigvalueWishart} and has been tied to the Tracy-Widom distribution. A similar analysis has been performed for complex Gaussian channels \cite{Kazakopoulos2011_LivingAtTheEdge_LD_MIMO}.

\subsection{Discussion}
\label{sec:Optical_discussion}

In this section we briefly discuss limitations of the above results.

One possible criticism of the above analysis may be whether the expression of the mutual information used represents the true transmission rate for optical MIMO channels. After all, this expression assumes a complex Gaussian input signal, which at this point does not correspond to what is used in current optical communications systems. Nevertheless, complex Gaussian input is optimal when the noise is also Gaussian, as it happens to be \cite{Cover_Thomas_book}. In addition, current modulation techniques are currently not too far from the ones used in wireless communications systems and therefore the above results can be taken as a figure of merit for the optical channel.

Other limitations of the above methodology have to do with the channel model used. For example, the fiber-optical channel has non-uniform mixing between different modes, as well as mode-dependent loss, in which the attenuation of each channel is different and in fact random \cite{Winzer2011_OpticalMIMOCapacity}. Such details can in principle be included in the channel, within the current random matrix framework, by appropriately generalizing the statistics of the random matrix.

Another important limitation of this analysis is the omission of non-linearities, which are inherently present in fiber optical communications, especially for large distance light propagation. Although some models have been applied in this context for single mode fibers \cite{Mitra_Stark_2001}, a coherent approach for the capacity of the non-linear optical MIMO channel is currently missing. However, it is hoped that the large number of channels will provide a small parameter for meaning approximations in the problem.

\section{Conclusions and Outlook}
\label{sec:Conclusions}

Modern telecommunications systems and algorithms are becoming increasingly complex and there is a need for mathematical tools that can tackle this complexity. Random matrix theory continues to be successful, not only in providing answers relevant in design and performance predictions, but also in providing intuition on the relevant issues. It is hoped that this introduction to the applicability of random matrix theory in communications and the description of how it can be  used to tackle real problems will further inspire the cross-fertilization between these fields.

It is worth mentioning that tools developed  in spin-glasses have also seen many applications in communications, signal processing and optimization. Nevertheless, both methodologies are essentially mean-field based. Perhaps the ``last'' frontier for physics applications in telecommunications will be in the description of spatial and temporal fluctuations in 2-dimensional wireless networks, where mean-field approaches do not hold.


%

\end{document}